\documentclass[aps,twocolumn,pre,superscriptaddress, floatfix]{revtex4-2}

\usepackage{xcolor,hyperref}
\hypersetup{
   colorlinks,
   linkcolor={blue!50!black},
   citecolor={blue!50!black},
   urlcolor={blue!80!black}
}

\usepackage{amsmath}
\usepackage{amssymb}
\usepackage{color}

\usepackage{graphicx}

\DeclareMathAlphabet{\mathitbf}{OML}{cmm}{b}{it}

\DeclareMathAlphabet{\mathsfit}{T1}{\sfdefault}{\mddefault}{\sldefault}
\SetMathAlphabet{\mathsfit}{bold}{T1}{\sfdefault}{\bfdefault}{\sldefault}

\newcommand{\zerovector}{\mathBold 0}
\newcommand{\ket}[1]{|#1\rangle}

\newcommand{\xv}{\mathitbf x}

\newcommand{\calBold}[1]{\mbox{\boldmath${\cal #1}$}}
\newcommand{\mathBold}[1]{\mbox{\boldmath$#1$}}
\newcommand{\dbar}{{\,\mathchar'26\mkern-12mu d}}

\begin{document}

\title{\fontsize{11}{3} \selectfont Note: Simple argument for emergent anisotropic stress correlations in disordered solids}

\author{Edan Lerner}
\email{e.lerner@uva.nl}
\affiliation{Institute for Theoretical Physics, University of Amsterdam, Science Park 904, Amsterdam, Netherlands}

\maketitle

It is now well-established that mechanical equilibrium in athermal disordered solids gives rise to anisotropic spatial correlations of the coarse-grained stress field \cite{Bulbul_pre_2009,Harrowell_jcp_stress_correlations_2016,lemaitre_pre_2017_stress_correlations_2D,lemaitre_jcp_2018_stress_correlations_3D,eric_d_field_theory_prl_2018,eric_d_field_theory_pre_2018,eric_Shimada_field_theory_arXiv_2020,tanaka_emergent_solidity_nat_comm_2020,Bulbul_prl_2020_emergent_elasticity} that decay in space as $1/r^{\dbar}$, where $r$ is the distance from the origin, and $\dbar$ denotes the spatial dimension. In this note we present a simple, geometry based argument for the scaling form of the emergent spatial correlations of the stress field in disordered solids. The presented approach bears some conceptual similarities with the field-theoretic approach of \cite{eric_d_field_theory_prl_2018,eric_d_field_theory_pre_2018,eric_Shimada_field_theory_arXiv_2020}. 

Consider a disordered solid whose constituent particles interact via pairwise, radially-symmetric interactions. We denote particle coordinates by $\xv_i$, the radius vector between a pair $i,j$ of interacting particles by $\xv_{ij}\!\equiv\!\xv_j\!-\!\xv_i$, and the distance between them by $r_{ij}\!\equiv\!\sqrt{\xv_{ij}\!\cdot\!\xv_{ij}}$. With these definitions, the magnitude of the forces $f_{ij}$ between particle $i$ and particle $j$ are defined to be positive (negative) if they are repulsive (attractive), and the condition of mechanical equilibrium reads
\begin{equation}\label{foo00}
\calBold{F}_i = \sum_j f_{ij} \xv_{ij}/r_{ij} = \zerovector\,,
\end{equation}
where $\calBold{F}_i$ denotes the \emph{net} force on the $i$'th particle.

Eq.~\eqref{foo00} can be cast into a more convenient, bra-ket notation in the form \cite{calladine_1978,asm_pnas_2012,states_of_self_stress_epje_2018}
\begin{equation}
\ket{\calBold{F}} = {\cal S}^T\ket{f} = \zerovector\,,
\end{equation}
where we have defined the linear operator ${\cal S}$ with components
\begin{equation}\label{foo04}
{\cal S}_{ij,k} \equiv \frac{\partial r_{ij}}{\partial\xv_k} = (\delta_{jk} - \delta_{ik})\frac{\xv_{ij}}{r_{ij}}\,,
\end{equation}
which takes vectors from the space of particles' coordinates, to the space of pairs of interacting particles ($\delta_{jk}$ is the Kronecker delta). ${\cal S}$ is constructed based on \emph{geometric} information of the network of pairwise interactions of a disordered configuration. States occupying the null-space of ${\cal S}^T$ --- which result in zero net force $\calBold{F}_i\!=\!\zerovector$ on each particle --- are known as \emph{states of self stress} \cite{vanSaarloos_prl_2004,sussman_soft_matter_2016,states_of_self_stress_epje_2018}. The number of orthogonal states of self stress of a given configuration scales as $z\!-\!2\dbar$ where $z$ is the mean coordination of the network of interactions \cite{sussman_soft_matter_2016}.

Consider next a disordered solid configuration, and assign any set of \emph{random} forces $\phi_\alpha$ between the particles, which obey the statistics
\begin{equation}\label{foo01}
\overline{\phi_\alpha} = 0\,, \quad\mbox{and}\quad \overline{\phi_\alpha\phi_\beta} = \delta_{\alpha\beta}\,,
\end{equation}
where here and in what follows Greek letters label \emph{pairs} of interacting particles, $\overline{\bullet}$ denotes a spatial average, $\delta_{\alpha\beta}\!=\!1$ if the pair $\alpha$ is equal to the pair $\beta$, and $\delta_{\alpha\beta}\!=\!0$ otherwise. The assigned `ancestral' forces $\phi_\alpha$ are supposed to represent the random forces between particles of a liquid before vitrification, shown recently \cite{tanaka_emergent_solidity_nat_comm_2020} to be uncorrelated in liquids residing just above their respective glass transition temperatures. Since the forces $\phi_\alpha$ are random, then clearly ${\cal S}^T\ket{\phi}\!\ne\!\zerovector$, i.e.~the system is out of mechanical equilibrium. We note that the $\delta$-correlated ancestral forces requirement is analogous to the integrability condition on the pressure autocorrelation as discussed in Refs.~\cite{Bulbul_pre_2009,lemaitre_pre_2017_stress_correlations_2D,lemaitre_jcp_2018_stress_correlations_3D,eric_d_field_theory_prl_2018,eric_d_field_theory_pre_2018,eric_Shimada_field_theory_arXiv_2020}.

We now project the set of random forces $\ket{\phi}$ onto the null space of ${\cal S}^T$, to obtain a state of self stress $\ket{f}$, namely
\begin{equation}
\ket{f} = {\cal P}\ket{\phi}\,, \quad\mbox{such that}\quad {\cal S}^T\ket{f}=\zerovector\,.
\end{equation}
As mentioned above regarding ${\cal S}$, the projection operator ${\cal P}$ is also constructed using geometric information of the network of interactions alone; it is given by \cite{matthieu_thesis,states_of_self_stress_epje_2018,comment_sss_soft_matter_2017}
\begin{equation}\label{foo03}
{\cal P} \equiv {\cal I} - {\cal S}\big({\cal S}^T{\cal S}\big)^{-1}{\cal S}^T\,,
\end{equation}
where ${\cal I}_{\alpha\beta}\!\equiv\!\delta_{\alpha\beta}$ is the interacting-pairs-space identity operator. It is easy to show that ${\cal P}\!=\!{\cal P}^T\! =\! {\cal P}^2$. ${\cal P}$ is, in essence, the discrete analog of the ``transverse projector" operator described in \cite{eric_d_field_theory_prl_2018,eric_d_field_theory_pre_2018}. 

Finally, consider the spatial force correlations
\begin{equation}
\overline{f_\alpha f_\beta} = \overline{\phi_{\kappa}{\cal P}_{\kappa\alpha}^T{\cal P}_{\beta\chi}\phi_{\chi}}\,,
\end{equation}
where repeated indices are understood as summed over. Since the random ancestral forces $\phi_\alpha$ are uncorrelated with the interaction network's geometry encoded in ${\cal P}$, then $\overline{\phi_{\kappa}{\cal P}_{\kappa\alpha}^T{\cal P}_{\beta\chi}\phi_{\chi}}=\overline{{\cal P}_{\kappa\alpha}^T{\cal P}_{\beta\chi}}\,\overline{\phi_{\chi}\phi_{\kappa}}$. Using Eq.~(\ref{foo01}), we find
\begin{equation}\label{foo02}
\overline{f_\alpha f_\beta} = \overline{{\cal P}_{\kappa\alpha}^T{\cal P}_{\beta\chi}}\delta_{\kappa\chi} = \overline{{\cal P}_{\alpha\beta}}\,.
\end{equation}
We point out that (i) the elastic Green's function of a relaxed Hookean spring network derived from any solid's network of pairwise interactions is given by $({\cal S}^T{\cal S})^{-1}$ (assuming unit stiffnesses and masses), as shown in \cite{asm_pnas_2012,states_of_self_stress_epje_2018}. (ii)~${\cal P}_{\alpha\beta}$ (with $\alpha\!\ne\!\beta$) is precisely the displacement response --- projected onto the radius vector of the $\beta$ pair (see definition above) --- that results from applying a unit force dipole on the pair $\alpha$ of the corresponding mapped elastic Hookean spring network \cite{states_of_self_stress_epje_2018}.

According to continuum elasticity theory, the displacement response to a force dipole decays as $1/r^{\dbar-1}$ at large distances $r$ from the force dipole. The average projection of such a displacement response onto the radius vector of an interacting pair located away from the applied force dipole is akin to taking the displacement-response's spatial gradient (due to the difference $\delta_{jk}\!-\!\delta_{ik}$ as appears in Eq.~(\ref{foo04}), see also~\cite{breakdown, SciPost2016}). Therefore $\overline{f_\alpha f_\beta}\!\sim\!\overline{{\cal P}_{\alpha\beta}}\!\sim\! 1/r^{\dbar}$, in agreement with previous work regarding stress correlations in disordered solids~\cite{Bulbul_pre_2009,Harrowell_jcp_stress_correlations_2016,lemaitre_pre_2017_stress_correlations_2D,lemaitre_jcp_2018_stress_correlations_3D,eric_d_field_theory_prl_2018,eric_d_field_theory_pre_2018,tanaka_emergent_solidity_nat_comm_2020,Bulbul_prl_2020_emergent_elasticity,eric_Shimada_field_theory_arXiv_2020}. We note that the anisotropy seen in far-field stress correlations ($\sim\!\cos(4\theta)$ in two-dimensions, with $\theta$ the azimuthal angle) can also be obtained via the aformentioned analogy to the elastic Green's function and the continuum-elastic response to local force dipoles. 

Interestingly, the same exercise can be repeated assuming random ancestral forces $\phi_\alpha$ in a periodic lattice of rigid struts. In that case, one finds $\overline{f_\alpha f_\beta}\!\sim\! {\cal P}_{\alpha\beta}\!\sim\! r^{-\dbar}$ (no spatial average over ${\cal P}$), where ${\cal P}$ is exactly obtained via Eq.~(\ref{foo03}), given any lattice geometry. This means that \emph{structural} disorder itself --- and therefore also isotropy --- are not necessary ingredients for the form of emergent spatial correlations of coarse-grained stress fields.

We thank Gustavo D\"uring, Eran Bouchbinder and Eric DeGiuli for discussions. Financial support from the NWO (Vidi grant no.~680-47-554/3259) is acknowledged. 

\section*{Data availability}
Data sharing is not applicable to this article as no new data were created or analyzed in this study.

\end{document}